# A chemist's view of chemical bonding in the mechanism of high temperature superconductivity

Michel Pouchard*, Antoine Villesuzanne* and Alain Demourgues

CNRS, ICMCB, UPR 9048, F- 33600 Pessac, France

Univ. Bordeaux, ICMCB, UPR 9048, F-33600 Pessac, France

*(In memory of Paul Hagenmuller 1921-2017)*

*michel.pouchard@icmcb.cnrs.fr ; antoine.villesuzanne@icmcb.cnrs.fr

**Abstract**

Through a 2-dimensionnal tight-binding crystal orbital approach and a $(CuO_2)_2$ square unit cell of parameter $a$, we show that a $Cu^{2+}/O^{2-} \rightarrow Cu^{+}/O^{-}$ charge transfer is likely to occur at the $M(\pi/a, \pi/a)$ point of the Brillouin zone, for $O_4$ groups with antibonding $b_{1g}$ symmetry. This approach emphasises the role of oxygen-oxygen interactions in avoiding the nesting of the Fermi surface and agrees with its observed topology. At the M point of the Brillouin zone, oxygen atoms are strongly dissymmetric ("Janus" atoms) and link copper atoms with different environments ($a_{1g}$ vs. $b_{1g}$ symmetries). Further hole doping generates two situations: two holes (S=0) and/or a single hole (S=+/- ½) in the $O_4$ $b_{1g}$ groups ($\sigma$ or $\pi$), with a possible equilibrium between them; the former can be considered as a "hole lone pair" by analogy with electron lone pairs. Mulliken-Jaffe electronegativity considerations justify the nearly zero U Hubbard parameter value. The mixing of the pair-occupied with pair-unoccupied wave functions is realised via an electronic Hamiltonian in place of the electron-phonon-coupling of the pristine BCS theory.

# I. INTRODUCTION

As mentioned as early as 1990 by Sir Nevill Mott, "to understand the new superconductors, it is essential to bring together the insights of chemistry and physics" [1] (Sentence related to the pioneering works of P. Hagenmuller and his co-workers on +III perovskite cuprates). Here, we present a tentative scenario for some key aspects in high-$T_C$ superconductors, based on arguments familiar to solid state chemists but yet emerging from solid state physics. Crystal structure and translation symmetry, via reciprocal space (or k-space, k for momentum), bring specific and considerable constraints to (real-space) chemical bonding and electronic states. Examining those constraints in detail for $CuO_2$ layers of cuprate superconductors, especially at Fermi level, is a simple but fruitful, probably prerequisite approach of the building blocks for a comprehensive mechanism of high-$T_C$ superconductivity.

For a review of the state-of-the-art on high-$T_C$ superconductors (HTSC), the reader will advantageously refer to papers in Refs. 2-5. Here, we also present recent developments of our previous works on so-called optimally- and over-doped compositions of high-$T_C$ cuprates,[6] consistent with the Bardeen, Cooper and Schrieffer (BCS) theory.[7]

# II. A STEP-BY-STEP TIGHT-BINDING APPROACH OF CHEMICAL BONDING AND BAND STRUCTURE OF A $(CuO_2)^{n-}$ LAYER

The pseudo-gap part of the HTSC phase diagram (under-doped compositions) has been extensively studied in recent years. In the range from 5 to 12 % hole doping rate, it presents many phases above the critical temperature $T_C$: strange metal, charge density waves (CDW) with stripe or checkerboard topologies, pair density waves (PDW). Many works have been devoted to the possible links between those phases and the superconductivity mechanism in optimally-doped HTSC.

Our approach is opposite: starting from the Fermi liquid state, we introduce progressively a more localised picture, via the consideration of i) molecular orbitals in the unit cell (intra-cell interactions in real space), then ii) inter-cell interactions and translation symmetry (momentum space).

The tight-binding parameters are the interatomic overlap $S_{ij} = \langle \chi_i | \chi_j \rangle$, the exchange (or transfer) integral $t_{ij} = \langle \chi_i | H^{eff} | \chi_j \rangle$ and the Coulomb integrals $H_{ii} = \langle \chi_i | H^{eff} | \chi_i \rangle$, where $| \chi_j \rangle$ are the atomic orbitals (AO) and $H^{eff}$ is the effective Hamiltonian operator. Both Cu-O and O-O interactions have been considered. In the text, Cu-O(2p) transfer integrals are noted $t_s$, $t_p$ and $t_d$ with respect to the Cu AO involved, and O-O transfer integrals are noted $t_{OO}$. We performed explicit extended Hückel tight-binding (EHTB) calculations,[8,9] i.e. including overlaps for a full basis set of O (2s, 2p) and Cu (3d, 4s, 4p) AOs. In the schematic representations of relevant crystal orbitals, we have restricted the basis set to copper $3d_{x^2-y^2}$, 4s, $4p_x$ and $4p_y$ AOs, and oxygen $2p_x$ and $2p_y$ AOs.

### A. Intra-cell $(Cu_2O_4)^{-n}$ molecular orbitals.

We consider here the square planar unit cell of composition $(Cu_2O_4)^{n-}$ and cell parameter $a = a' \cdot \sqrt{2}$, where $a'$ is the cell parameter for the primitive $CuO_2$ layer. Fig. 1(a) shows these two unit cells and their corresponding Brillouin zones (BZ), with high symmetry points labelled $\Gamma(0;0)$, $X(\pi/a;0)$, $Y(0;\pi/a)$ and $M(\pi/a;\pi/a)$ (Fig. 1(b)).

In a previous tight-binding investigation of the $La_4LiCu^{3+}O_8$ oxide, we had shown that O-O interactions were large, with $t_{OO}$ values of a few hundred meV and that a charge transfer $Cu^{3+} + 2O^{2-} \rightarrow Cu^{+} + 2O^{-}$ occurred.[10,11] Here, it is thus sound to build up first the oxygen MOs in the unit cell and the corresponding band structure. Copper AOs will be accounted for in a second stage, according to the symmetry of oxygen crystal orbitals, in the same way Zheng and Hoffmann described the electronic structure of $BaMn_2P_2$.[12]

Fig. 1c presents the eight 2-dimensional oxygen MOs in the $Cu_2O_4$ unit cell, labelled by σ or π, bonding (B), non-bonding (NB) or antibonding (AB), and *gerade* (g) or *ungerade* (u) symmetry.[13] Note that the hybridization of σ-$1e_u$ and π-$2e_u$ is symmetry-allowed. Fig. 1(d) illustrates the cell-to-cell phase relation as a function of cell coordinates (m and n) and wave vector $k(k_x;k_y)$, according to $\Psi_k \propto \Sigma_m \Sigma_n \exp(ik_x ma) \cdot \exp(ik_y na) \cdot \chi_{(r)}$, where a is the cell parameter and $\chi_{(r)}$ is the intra-cell expansion of the wave function (MO).

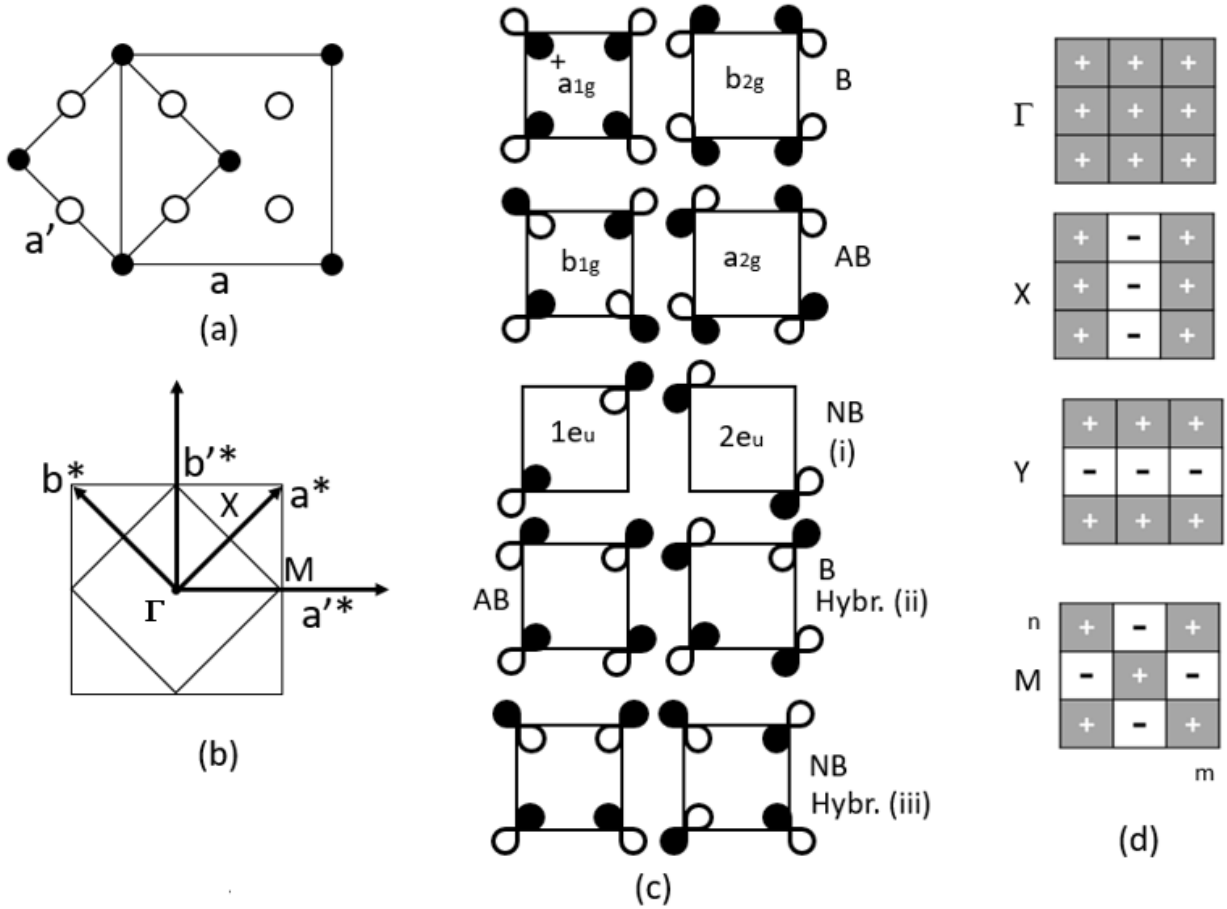

FIG. 1. (a) Unit cells and (b) Brillouin zones for two-dimensional $CuO_2$ and $Cu_2O_4$ layers. (c) Schematic representation of symmetry-adapted molecular orbitals in the unit cell, considering oxygen only. A, B and NB refer to antibonding, bonding and non-bonding orbitals, respectively. (i), (ii) and (iii) refer to transfer integral values in Table 1. (d) Phase factors for Bloch functions, for several wave vectors in the Brillouin zone, and unit cells labelled m and n.

Table I. Transfer integral terms for the upper oxygen bands at $\Gamma$, X and M points of the BZ. (i), (ii) and (iii) refer to orbital representations in Fig. 1(c).

|  | $\Gamma$ | X | M |
|---|---|---|---|
| σ-a1g | $4t_{OO} + 2t_p$ | $2t_{OO} + 2t_p$ | $t_p - t_d$ |
| σ-b1g | $-4t_{OO} - 2t_d$ | $-2t_{OO} + t_p - t_d$ | $t_p - t_d$ |
| π-a2g | $-4t_{OO}$ | $-2t_{OO} + 0.5t_p$ | 0 |
| π-b2g | $+4t_{OO}$ | $2t_{OO} + 0.5t_p$ | 0 |
| σ/π-2eu | $2t_p$ (ii) | $-2t_{OO} + t_p - t_d$ (iii) | $t_p - t_d$ |
|  |  | $2t_{OO} + 2t_p$ (iii) | $2t_p$ |

At this stage, two remarks can be made: i) a pure oxygen framework presents here the same local symmetry than the transition-metal 3d AOs, as exemplified in Fig. 2(a), where $a_{2g}$ and $b_{1g}$ symmetry are those of the copper unoccupied and occupied sites, respectively; ii) the eight oxygen bands are degenerate and overall non-bonding at the M point of the BZ.

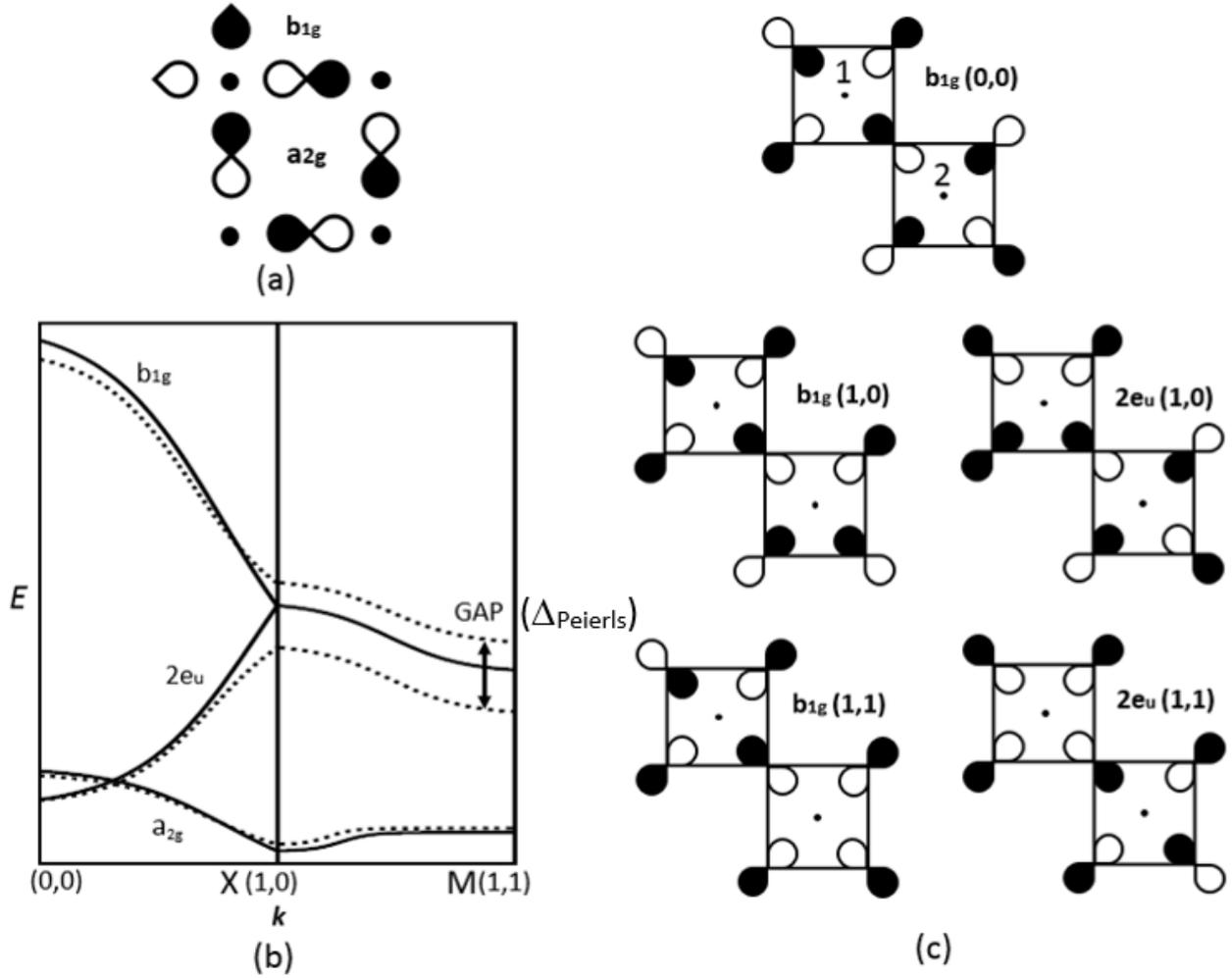

FIG. 2. (a) d-type symmetry of oxygen molecular orbitals, at copper occupied and vacant sites. (b) Dispersion curves E(k) for a $Cu_2O_4$ layer with (dotted lines) and without $a_{1g}$-type breathing distortion, around Fermi level, calculated by the EHTB method. (c) Crystal orbital mapping of oxygen wave functions of the two most important bands, σ-$2e_u$ and σ-$b_{1g}$ in Γ, X and M respectively, with their degeneracy and similitude by permutation of copper in sites 1 and 2.

B. On the role of Cu-O transfer integrals $t_s$, $t_p$ and $t_d$.

In a second stage, we examine the mixing of copper AOs with oxygen crystal orbitals, as a function of symmetry and energy difference factors. Two situations occur that differentiate copper 4s and 4p AOs from copper 3d AOs:

(i) Cu 4p or 4s (which is generally hybridized with $3d_{z^2}$) orbital energies lie well above the energy range of oxygen 2p blocs. Here, the 4p/4s-oxygen mixing will overall stabilize the oxygen 2p bloc and destabilize 4p- and 4s-character states.

(ii) The situation for $3d_{x^2-y^2}$ is very different since Cu 3d and O 2p orbitals are almost degenerate. According to Ref. 8, O 2p orbitals should lie less than 1 eV above Cu 3d orbitals, a difference that is furthermore affected in sign and magnitude by ligand field and Madelung potential, among other factors. According to this quasi-degeneracy or slightly higher energy of 2p vs. 3d orbitals, it is reasonable to consider a charge transfer from oxygen to copper, formally written $Cu^{2+} + O^{2-} \rightarrow Cu^{+} + O^{-}$ implying a mixed valence state for O (mean oxidation state of –1.50). Another consequence of this quasi-degeneracy is an expected low value of the Hubbard U parameter (proportional to the Racah B parameter), allowing a one-electron band theory description.

With such choices, the oxygen crystal orbital bloc tends to be stabilised by mixing with Cu 4s and 4p and, conversely, destabilised by mixing with Cu $3d_{x^2-y^2}$. Due to the 2p-3d quasi-degeneracy, the latter effect should dominate, corresponding to $t_p \approx t_s \ll t_d$.

Fig. 2b shows the dispersion curves E(k) after copper-oxygen mixing, for the top of the oxygen band bloc, calculated by the EHTB method. Table I gives the corresponding tight-binding transfer terms for these bands at high-symmetry points of the BZ. In increasing energy values, the band structure is composed of:

- a doubly occupied σ-$b_{1g}$ Cu-character band, purely of copper character at Γ, then progressively involving oxygen crystal orbitals of the same $b_{1g}$ symmetry, in a Cu-O bonding manner and illustrating the charge transfer situation,
- four other doubly occupied Cu 3d AOs, weakly- or non-dispersive,
- a highly dispersive O-character bloc.

Note that: i) σ/π-$2e_u$ and σ-$b_{1g}$ bands are degenerate between X and M, ii) π-$a_{2g}$ and π-$b_{2g}$ bands (π-$b_{2g}$ is not represented in Fig. 2(b)) are purely of oxygen character at Γ as well as at M, and degenerate at M.

### C. Symmetry at the copper sites of the oxygen band bloc.

Fig. 2(c) (top) gives a schematic representation of the σ-$b_{1g}$ crystal orbital at Γ, and of σ/π-$2e_u$ and σ-$b_{1g}$ crystal orbitals at X and M points of the BZ. At Γ, both bands present copper sites with their own symmetry only ($e_u$ and $b_{1g}$). However, at X and M, both bands present a mixed situation, with sites of $e_u + b_{1g}$ and $a_{1g} + b_{1g}$ symmetries in 1:1 ratio, explaining their degeneracy. Note the succession of mirror planes (m) and anti-symmetric planes (m−) along the a and b axis, for both bands at M.

### III. ELECTRON COUNTING AND FERMI LEVEL

For the n-type cuprates with a $3d^{9+\delta}$ electron configuration, the Fermi level ($E_F$) crosses the σ-$2e_u$ and σ-$b_{1g}$ bands between Γ and X, forming a hole-pocket centred at X, whose extension increases with hole doping rate (Fig. 2(b) and 3). For a hole concentration of approximately 10%, $E_F$ reaches the two degenerate bands σ-$2e_u$ and σ-$b_{1g}$ in M. Interestingly, this hole concentration value is close to the pseudo-gap phase limit for the p-type cuprates.

As noticed by Whangbo *et al.* in many HTSC series,[14] the optimal doping does not generally correspond to a unique hole concentration neither to a unique Cu-O bond length. Our model is able to explain this anomaly by the shape and the extension of the Fermi surface (σ-$2e_u$ band between Γ and X), which both depend of the magnitude of the band overlap at X and M points. This overlap, which give rise to the electron pocket at M, is driven both by $t_{OO}$ (Table I) and by the competing O-X bonds (X=$Ln^{3+}$, $AE^{2+}$).

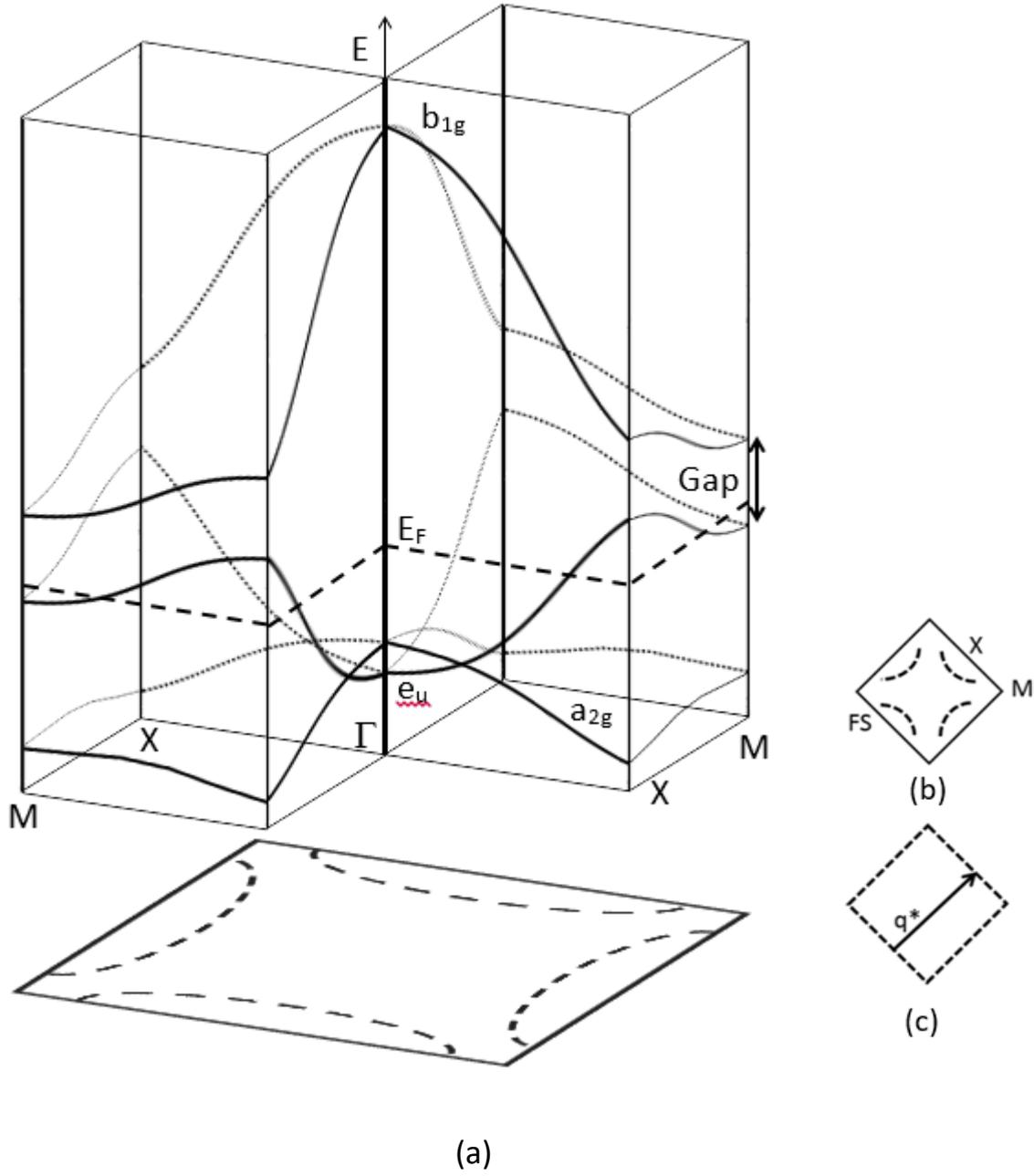

FIG. 3 (a) Dispersion curves for the σ-2$e_u$, σ-$b_{1g}$ and π-$a_{2g}$ bands near $E_F$, as in Fig. 2(b). Fermi surface (b) with and (c) without oxygen-oxygen interactions ($t_{OO}$). $t_{OO} \neq 0$ removes the Fermi surface nesting (q*) and contributes to the gap opening.

# IV. CHARGE ORDER FROM FERMI SURFACE NESTING, ELECTRON CORRELATIONS, OR ORBITAL/BOND ORDERING ONLY?

The first question to be addressed is whether a charge order state (CO) is also achievable in the optimal doping part of the phase diagram, as it may also compete with superconductivity (as well established for the pseudo-gap phase [2]).

## A. Importance of the $t_{OO}$ resonance integral.

As seen in Table I, $t_{OO}$ contributes significantly to the width of the oxygen band stack, by $8t_{O-O}$ between σ-$a_{1g}$ and σ-$b_{1g}$ at Γ. It is clear that vanishing O-O interactions ($t_{OO} = 0$) would lead to non-dispersive oxygen bands along the X-M path in the BZ, and to a square Fermi surface such as the one for a simple half-filled $3d_{x^2-y^2}$ band, with a perfect nesting of vector $q^* = ½(a'^* + b'^*)$ (Fig. 3(c)). Therefore, O-O interactions are essential for avoiding a charge ordering of charge-density wave type.

## B. Which types of order?

From Fig. 2(c), it appears that copper sites labelled 1 and 2, for the M point and the σ-$2e_u$ and σ-$b_{1g}$ bands, are differentiated by i) symmetry: $a_{1g}$ vs. $b_{1g}$, as previously quoted, ii) sign of O-O overlap, bonding vs. antibonding for $a_{1g}$ and $b_{1g}$, respectively, iii) Cu-O interactions type and sign: weakly bonding for $a_{1g}$ with Cu $4s/3d_{z^2}$ involved (no nodal surface around copper); strongly antibonding for $b_{1g}$ with Cu $3d_{x^2-y^2}$ involved (nodal planes between Cu and O).

Such critical differences between $a_{1g}$ and $b_{1g}$ sites should result in charge ordering, to be rather named as bond order (BO) or possibly orbital order (OO), and which involve the whole framework.

Along Cu-O directions, oxygen thus behaves as a "Janus atom"; such a duality together with the high oxygen polarizability should result in large electric dipoles $\mu_i$ at oxygen sites (Fig. 4(a)). Several situations can be subsequently envisaged:

(i) a Peierls-type static distortion with shortened vs. lengthened Cu-O bonds at $a_{1g}$ and $b_{1g}$ sites, respectively, lifting the degeneracy of σ-$2e_u$ and σ-$b_{1g}$ bands (Figs. 2(b) and 3). From our tight-binding calculations, the ratio of gap magnitude vs. atomic displacement is ca. 50 meV/pm;

(ii) above a critical temperature $T_{BO}$, oxygen atoms may oscillate along the Cu-O directions, according to a breathing phonon mode (shown by an arrow in Fig. 4(b)), leading to a modulated gap (Fig. 4(c)),

(iii) during the timespan of vanishing oxygen displacement (each half-period), electrical dipole oscillations should take over phonon breathing, at a much higher frequency (Fig. 4(d)).

Situation (i) can be compared to that of $La_4LiCuO_8$, where the $[O_4]^{6-}$ group hosts two charge transfer holes in the $\sigma^*$-$b_{1g}$ molecular orbital.[11]

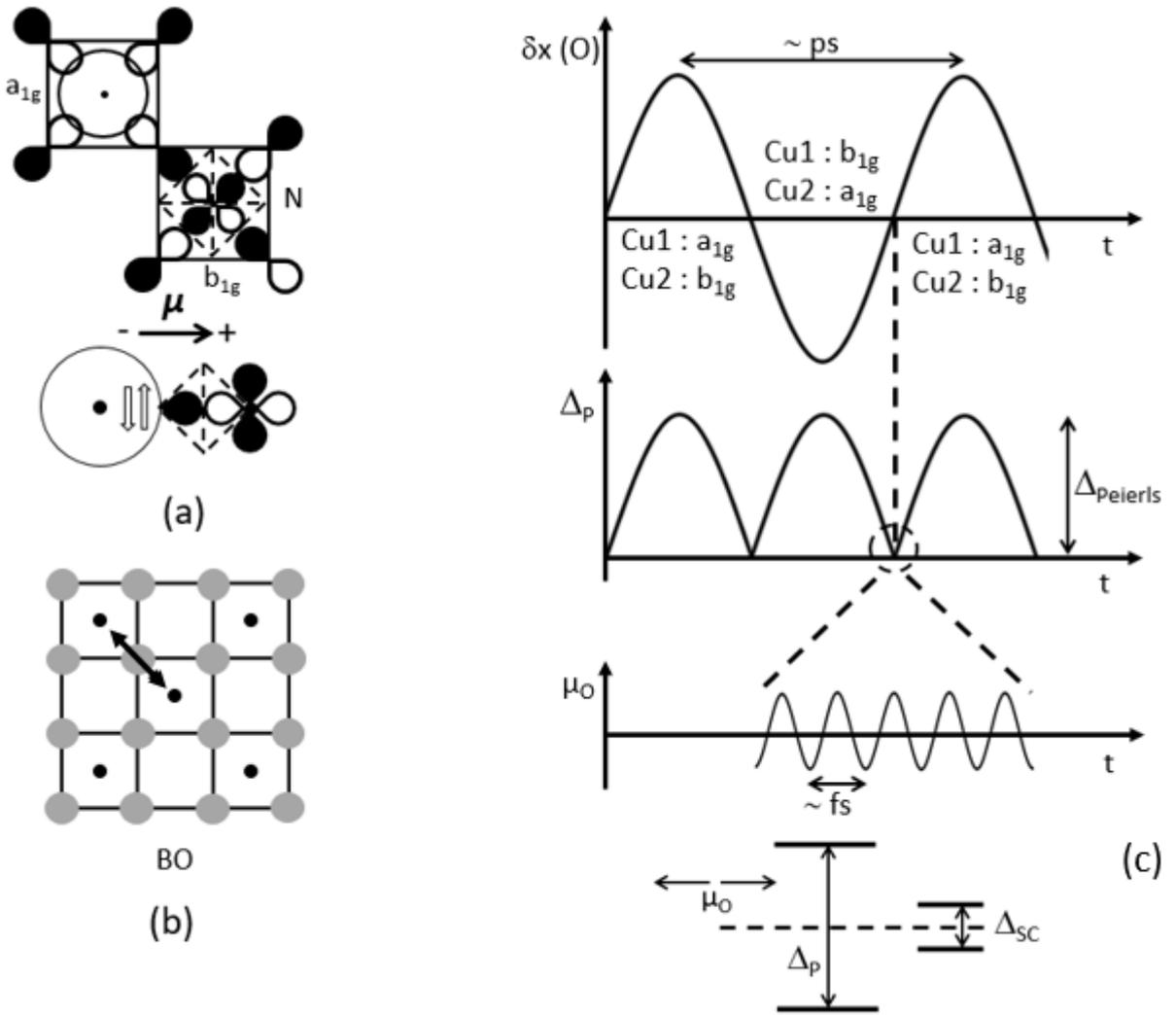

FIG. 4. (a) Strong differentiation of O-O and Cu-O bonding between $a_{1g}$ and $b_{1g}$ sites, with nodal surfaces (dashed lines) for the latter, inducing a high anisotropy (Janus-type) for the oxygen atoms and an electric dipole $\mu$. (b) Static or dynamical Peierls-type distortion associated to the oxygen anisotropy and displacement. (c) Top: oxygen atom displacement $\delta x = f(t)$ for the breathing phonon mode, switching the wave functions $a_{1g}$ and $b_{1g}$ on the two copper sites, with a picosecond timescale. Middle: Dynamical Peierls gap opening $\Delta_p$. Bottom: at each half-period (vanishing $\Delta p$ and distortion) and at a femtosecond timescale, oscillations of oxygen dipoles $\mu_O$ allow the state mixing in Eq. (2) leading to BCS-type superconductivity and gap ($\Delta_{SC}$).

## V. FURTHER DOPING AND NEGATIVE-U PAIRING

Two situations must be considered :

(1) $t_{OO} < (t_d - t_p)/4$ (see Table 1 and Fig. 2(b))

Hole doping at the top of the σ-2$e_u$ band will affect $b_{1g}$ sites only, i.e. half of the copper sites for $T < T_{BO}$. Two possible situations emerge for the distribution of doping holes (Fig. 4(a)):

(i) two holes per $b_{1g}$ site, for a $S = 0$ diamagnetic state, if the hole-hole Hubbard parameter $U_{h-h}$ is small compared to the $t_{OO}$ resonance integral. In addition, it can be shown that the Mulliken-Jaffe oxygen electronegativity $d(E_f)/dq$ (Fig.5) vanishes for a point charge of –0.60, close to the value of –1.00 expected from charge transfer ($O^{1.5-}$, i.e. $O_4^{6-}$) plus hole doping (2 hole per $Cu_2O_4$ formula unit, $O_4^{4-}$) (Fig. 4(a)),

(ii) one hole per $b_{1g}$ site ($S = ½$) state, with oxygen in the $O^{-1.25}$ state.

Both situations show very weak oxygen negative electronegativity ($d(E_f)/dq$) and comparable ionisation energy for $2\ O^{1.5-} \rightarrow 2\ O^- + e^-$ ($\Delta H = -5.9$ eV) and $4\ O^{1.5-} \rightarrow 4\ O^{1.25-} + e^-$ ($\Delta H = -6.7$ eV), corresponding to the equilibrium: $2\ O^{-1.5-} + 2\ O^- \rightarrow 4\ O^{1.25-}$ ($\Delta H = -0.8$ eV).

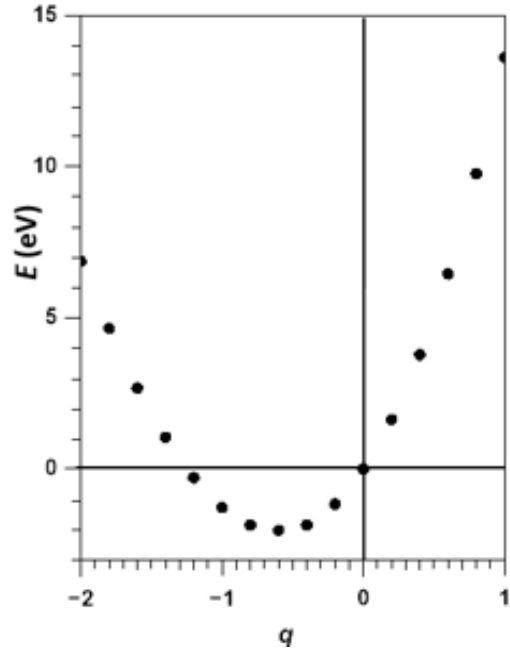

FIG. 5. Formation energy (eV) vs. charge q of an isolated $O^q$ ion (according to $E(q) = 6.925q + 6.190q^2 + 0.505q^3$). The derivative is the Mulliken-Jaffé electronegativity $\chi_{MJ}$, which is vanishing at $q = -0.6$ A negative value for the Hubbard U parameter may result from these very low electronegativity values, as well as from the very close formation energies for the entities $O_4^{4-}$ and $O_4^{5-}$ in next figure (E= -1.24 and +0.03 eV for q = -1 and -1.25, respectively).

The first case corresponds to a doping charge order (DCO), the second one to a p-type metal (Fig. 6(a)). An equilibrium between DCO and metallic states could be envisaged, favoring opposite spins in each $b_{1g}$ site, and analogous to a valence disproportionation according to class (III) of the Day and Robin classification for two electrons [15] (Fig. 6(b)) or, for the physicists, to the so-called "negative U pairing" situation. This type of equilibrium is exemplified by $Bi^{4+}$ in the superconductor $Ba_{(1-x)}K_xBiO_3$, where charge order and superconducting states compete,[16] or by the Tl-doped PbTe semiconductor where $Tl^{2+}$ disproportionates into $Tl^+$ and $Tl^{3+}$.[17] Here, hole lone pairs (HLP) are considered in place of the electron lone pairs involved in these two examples.

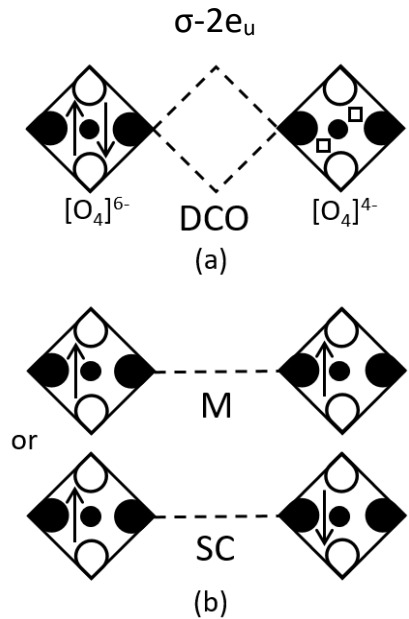

FIG. 6. Scheme of crystal orbitals along [100] for (a) a doping charge order (DCO) and (b) a metallic (M) or superconducting (SC) state, with one hole per site.

(2) For $t_{OO} \geq (t_d - t_p)/4$, an other type of hole pair localization can be envisaged, at the top of the $\pi$-$a_{2g}$ band overtaking the Fermi level in $\Gamma$, with several features:

- the $\pi$-$a_{2g}$ band is very flat (width ≈ 0.35 eV) as in molecular systems, which justifies our choice of a Mulliken-Jaffé approach (Fig. 5). Such a situation is consistent with previous observations that superconductivity in carbide halides as $Y_2B_2C_2$ is closely related to the occurrence of both steep bands and flat bands at Fermi level,[18]

- $\pi$-$a_{2g}$ is here purely oxygen in character at $\Gamma$-point (Table 1),

- it presents, at the non-occupied cationic position (centre of the a' single cell), a $b_{1g}$ symmetry site that is well adapted to this double hole doping; its positive Madelung potential is reduced by approximately 50% from the potential at the oxygen site. This $b_{1g}$ symmetry site is actually occupied in the rock salt-type layer of T-CuO.[19] The most antibonding part of the wave function can be represented as a large spot at that position (Fig 7(a)). Comparable spot-images, isolated or associated in pair or in files, have been observed by tunnelling-asymmetry imaging of samples of Na-doped $Ca_2CuO_2Cl_2$ (Na-CCOC) and Dy-doped $Bi_2Sr_2CaCu_2O_{8+\delta}$ (Bi-2212). (see for example Figs. 4B and 4E in Ref. 20).

For $T>T_{BO}$, the previous situations (see IV.B(iii)) do occur in an alternate manner; according to the phonon (or dipole oscillation) frequency.

## VI. A "JANUS" BCS-TYPE MECHANISM

The competition between CO – or CDW – and SC states has been a long time debate; already fifteen years ago, for example, Whangbo discussed the electronic instabilities of low-dimensional metals on the basis of band orbital mixing.[21] The electronic structure of a CDW, spin density wave (SDW) or SC can be described in terms of mixing between filled and empty states at the Fermi level of a normal metal. The perturbation at the origin of CDW and SC states is usually phonons, whereas the SDW instability arises from the on-site repulsion (Hubbard's parameter U). A key factor is the Fermi surface nesting, as for a half-filled $3d_{x^2-y^2}$ band, as mentioned above. Such a nesting can be partially destroyed by doping.

For CDW and SDW states, one-electron wave functions $\Phi(k)$ mix according to $<\Phi(k)|H'|\Phi(k')>$, where H' is a perturbation operator and k and k' are the occupied and unoccupied wave vector near the Fermi surface, respectively. For the SC state, the mixing terms involve an occupied pair function (a Cooper pair), $\Phi(k).\Phi(-k)$ and an unoccupied one according to:

$$<\Phi(k).\Phi(-k)|H'|\Phi(k').\Phi(-k')> \qquad (1)$$

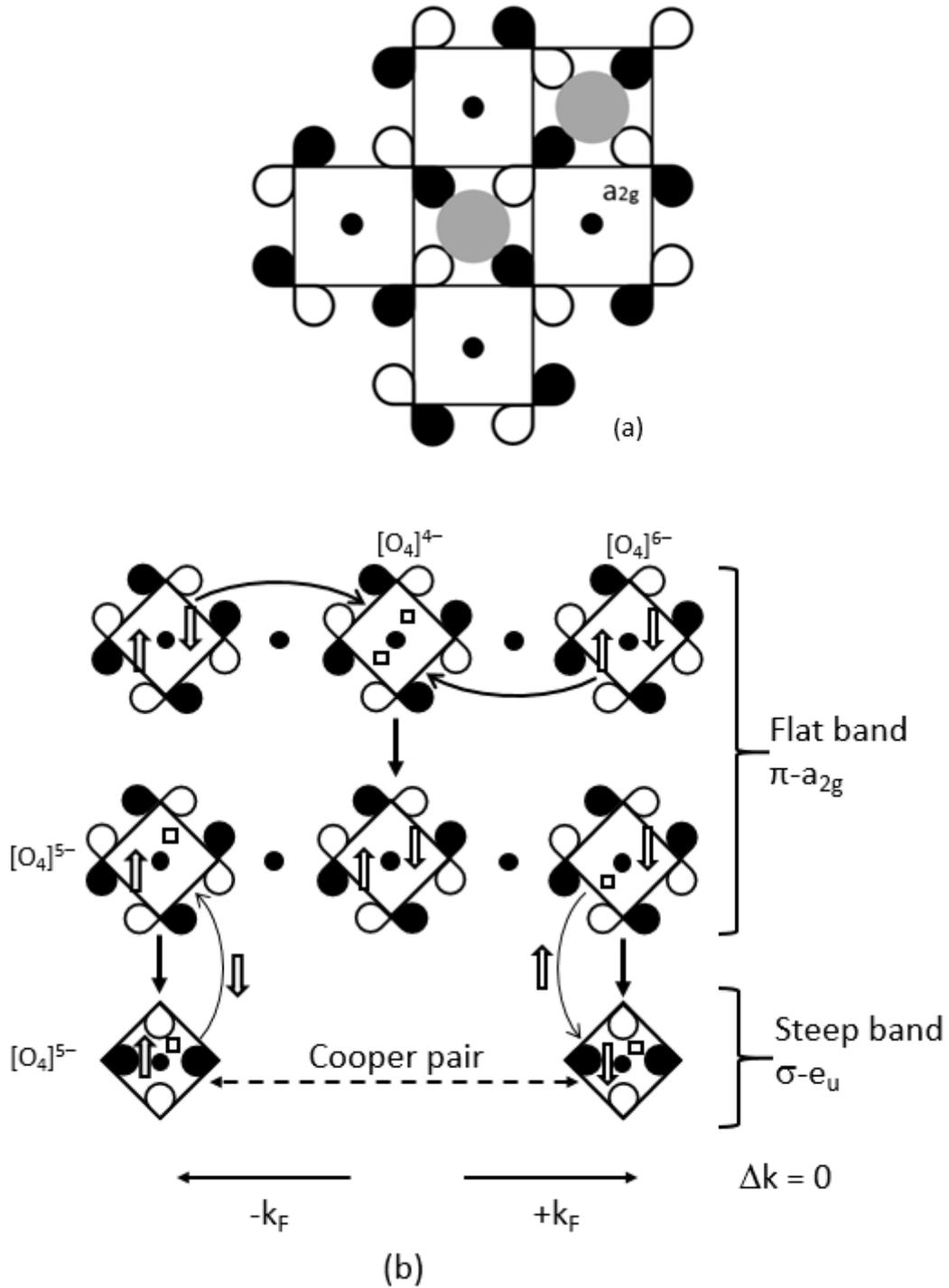

FIG. 7. (a) Double hole-doped $\pi$-$a_{2g}$ state, centered at the vacant cationic site of $b_{1g}$ symmetry (shaded area, corresponding to the most antibonding zone of this crystal orbital, and distributed similarly as spots observed in tunnelling-asymmetry imaging of Na-CCOC and Bi-2212 samples (Ref. 20). (b) Two hole $O_4^{4-}$ sites of the $\pi$-$a_{2g}$ flat band, in equilibrium with (top) filled $O_4^{6-}$ sites and (middle) a double number of singly-occupied $O_4^{5-}$ hole sites (S = ± ½ ) in the same band. These strongly polarized spins are coupled with those in the steep $\sigma$-$e_u$ band, forming Cooper pairs with S = 0 and $\Delta k$ = 0.

In the BCS model approach, H' is the electron-phonon coupling operator. It was concluded by M. Whangbo [21] that a large nesting of the Fermi surface should likely induce a metal-to-insulator transition of CDW or SDW type or, in some cases, a SC state if the fermion interactions are weaker than the bosonic ones. As shown in Section IV, the oxygen-oxygen interactions ($t_{OO}$) are amongst the most important parameters for avoiding the Fermi surface nesting and, consequently, promoting the SC state.

According to the depiction of crystal orbitals at Fermi level in the previous sections, Eq. (1) can be written more explicitly as:

$$<\Phi(\sigma\text{-}2e_u)(k_F).\Phi(\sigma\text{-}2e_u)(-k_F) \mid H' \mid \Phi(\sigma\text{-}b_{1g})(k_F').\Phi(\sigma\text{-}b_{1g})(-k_F')> \qquad (2)$$

What about H'? At a typical phonon frequency ($10^{-12}$ s), the system can be described as a succession of states with $a_{1g}$ and $b_{1g}$ symmetries, on each copper atom, in phase with the dynamic Peierls distortion. Beside, at the instant of nearly zero distortion, we can anticipate a mixing of these states at the frequency of an electric dipole oscillation. From a chemical point of view, such a mixing opens the superconducting gap. Eq. 2 is thus directly related to the negative U mechanism associating full and empty pair states.

We can thus write the class (III)-type equilibrium in the Kröger-Vink notation:

$O_4{''} + O_4{*}$ (S=0) $\leftrightarrow$ $O_4{^\cdot}$ (S=+1/2) + $O_4{^\cdot}$ (S=-1/2) ($\sigma$-$2e_u$) for the Cooper pair formation in the BCS model (Fig. 7).

## VII. AGREEMENT WITH RECENT EXPERIMENTAL DATA

This model can be discussed in regards to experimental facts in HTSC: [22]

- the upper $\sigma$-$b_{1g}$ band, mostly antibonding, is empty in agreement with the stability of the hole-doped HTSCs,
- the Fermi surface topology is correctly described by the tight-binding approach including $t_{OO}$,

- the bond order (below $T_{BO}$) and the doping charge order (DCO) are consistent with the 4a' characteristic length of the short range "stripe" or "checkerboard" order of the under-doped domain (Fig. 4),

- the mixing operator H', which induces a symmetry breaking, is of electronic nature as evidenced a few years ago by Dal Conte *et al.*[23] by time-resolved reflectivity in Bi-2212 samples, for "disentangling the electronic and phononic glue" in HTSCs.

In addition, energy-resolved local density of states (LDOS) imaging on $^{18}$O-enriched HTSC materials recently emphasized the role of a phonon of nearly 50 meV in the coupling mechanism;[2] here, we propose an indirect role for this phonon.

In support of our assumption of a BCS-type theory, it was shown by several techniques such as angle-resolved photoelectron spectroscopy (ARPES) and electron Raman scattering that the ratio $T_C$ vs. superconducting gap $\Delta_{SC}$ was nearly constant ($\approx$ 6) in several HTSC series, close to the value $\approx$ 4.3 predicted for a simple d-wave BCS gap (see, for example, Sacuto *et al.*[24]).

.As far as the symmetries of the order parameter and of the crystal orbital could be related, the $4s/3d_z^2$ hybridization would likely explain the mixed (d + s) character of the order parameter, as well as the inter-plane 3D coupling (6). This double contribution was discussed recently by Keller *et al.*,[25] who showed that a significant s-wave (20 to 25%) order coexists with a dominant d-wave gap in the $CuO_2$ layers. A predominant and robust d-form factor density wave, with approximately the same ratio, was also recently established by scanning tunnelling microscopy.[4]

## VIII. CONCLUSION

By examining the band structure of a $(Cu_2O_4)^{n-}$ layer, in particular at high-symmetry points of the Brillouin zone and at the vicinity of Fermi level, we have shown the critical role of oxygen-oxygen interactions on the topology of the Fermi surface, avoiding nesting that usually hinders superconductivity.

In a prominent charge-transfer situation, the two most important oxygen bands ($\sigma$-$2e_u$ and $\sigma$-$b_{1g}$) are strongly dispersive and non-degenerate at $\Gamma$, but degenerate at the M point of the Brillouin zone; both bands present the same succession of sites of $a_{1g}$ and $b_{1g}$ symmetries, creating a dissymmetric environment for all oxygen atoms in the layer. The low dispersion of these bands between X, Y and M points suggests a high density of states at the Fermi level. Our model explains the Fermi surface topology, with a superconducting gap near the X and Y points of the Brillouin zone for a $CuO_2$ layer.

We propose a dynamical bond order (BO) situation above a critical temperature $T_{BO}$ and/or a critical doping rate, below which a Peierls-type square distortion should freeze with a gap opening.

At all temperatures, further doping holes occupy only the $b_{1g}$ sites of the $\sigma$-$2e_u$ band, i.e. involving approximately half of the copper atoms; an equilibrium between states with one hole vs. two holes per site is proposed, especially in the $\pi$-$a_{2g}$ oxygen band with a possible doping charge order (DCO) and a negative-U pairing mechanism.

The mixing of occupied and unoccupied pair wave functions at Fermi level generates a symmetry breaking involving a new (SC) gap and the holes pairing (Cooper pairs), in the BCS theory framework, where the reversal of the oxygen polarization takes over the phonons role in the pristine BCS model.

Finally, this model may provide explanations for the joint role of electronic and phonon contributions to the pairing mechanism.

**Acknowledgements**

M. P. thanks Guy Deutscher and Andres Cano for fruitful discussions. Both authors thank J.-P. Doumerc for his contribution to the previous papers they devoted to high-$T_C$ superconductors.